\documentclass[11pt]{article}
\usepackage{appb,epsfig}

\def\SM{$\cal{SM}$}
\def\sw{s_W}
\def\cw{c_W}
\def\mW{M_W}

\def\mH{m_H}

\def\IJMP #1 #2 #3 {{\it Int.\ J.\ Mod.\ Phys.}\ {\bf #1}\ (#2) #3}
\def\MPL #1 #2 #3 {{\it Mod.\ Phys.\ Lett.}\ {\bf #1}\ (#2) #3}
\def\NC #1 #2 #3 {{\it Nuovo Cim.}\ {\bf #1} (#2) #3}
\def\NP #1 #2 #3 {{\it Nucl.\ Phys.}\ {\bf #1}\ (#2) #3}
\def\PL #1 #2 #3 {{\it Phys.\ Lett.}\ {\bf #1}\ (#2) #3}
\def\PR #1 #2 #3 {{\it Phys.\ Rev.}\ {\bf #1}\ (#2) #3}
\def\PP #1 #2 #3 {{\it Phys.\ Rep.}\ {\bf #1}\ (#2) #3}
\def\PRL #1 #2 #3 {{\it Phys.\ Rev.\ Lett.}\ {\bf #1}\ (#2) #3}
\def\RMP #1 #2 #3 {{\it Rev.\ Mod.\ Phys.}\ {\bf #1}\ (#2) #3}
\def\CMP #1 #2 #3 {{\it Comm.\ Math.\ Phys.}\ {\bf #1}\ (#2) #3}
\def\ZP #1 #2 #3 {{\it Z.\ Phys.}\ {\bf #1}\ (#2) #3}
\def\E #1 #2 #3 {{\bf #1}\ (#2) #3 (E)}

\date{July 6, 1998}
\begin{document}
\title{Two-loop large Higgs mass contribution to vector boson anomalous 
quartic couplings\thanks{Presented at Zeuthen Workshop on Loops and
Legs in Gauge Theories, Rheinsberg, Germany, 19-24 April, 1998.}
}
\author{G. Jikia
\address{Albert--Ludwigs--Universit\"{a}t Freiburg,\\
Fakult\"{a}t f\"{u}r Physik\\
Hermann--Herder Str.3, D-79104 Freiburg, Germany}
}
\maketitle
\begin{abstract}
The calculation of the two-loop corrections to the electroweak gauge
boson quartic couplings, growing quadratically with the Higgs boson
mass, is reviewed. The potential of the CERN Large Hadron Collider and
$e^+e^-$ linear collider to study such anomalous interactions is discussed.
\end{abstract}
  
\section{Introduction}
The remarkable precision of the electroweak experimental data
\cite{LP'97-1,LP'97-2} makes it possible to test the predictions of
the Standard Model (\SM) at the quantum loop level. After the
successful prediction of the top-quark mass from the $m_t^2$ one-loop
electroweak radiative corrections and the actual observation of the
top quark signal at the Tevatron, the mechanism of the spontaneous
electroweak symmetry breaking, connected to the existence of the Higgs
boson in the \SM{}, remains the last untested property of the
\SM{}. Electroweak observables are influenced also by the presence of
the Higgs boson, but contrary to the $m_t^2$ dependence at the
one-loop level they depend only logarithmically on the Higgs boson
mass. From the high-precision data at LEP, SLC and the Tevatron an
upper limit of $m_H<430$~GeV has been derived at the 95\% confidence
level \cite{LP'97-1,LP'97-2}. This bound is not very sharp however. In
a conservative conclusion the experimental limit may  be
interpreted in the \SM{} as an indication for a scale $m_H\leq {\cal
O}(1)$~TeV.

If the Higgs boson is really heavy the study of its indirect effects
at the quantum loop level at energies much smaller than $m_H$ will be
one of the most important goals for the future experiments.  Starting
from the two-loop level radiative corrections exhibit power growth for
$m_H\gg M_W$. Leading $m_H^2$ two-loop corrections to the
$\rho$-parameter \cite{rho,pisa} and to vector boson masses
\cite{masses} were calculated more than ten years ago.  These two-loop
large Higgs mass calculations were later extended to the case of the
triple vector boson couplings \cite{triple}. Power counting shows,
that only vertex functions with maximally four vector boson external
legs can have two-loop large Higgs mass corrections proportional to
$m_H^2$, while for five and higher point vertex functions no power
growth of the two-loop corrections with the Higgs mass is possible.
In this talk we review the results of our recent paper \cite{our},
where these calculations were completed and the analytical expressions
have been obtained for the two-loop $m_H^2$ corrections to quartic
electroweak gauge boson couplings in the \SM{} in the limit $m_H\gg
M_W$ at low energy $E\ll m_H$.

\section{The calculation}

In order to calculate the four vector boson
vertex function contribution to the low energy effective action
$\Gamma_{eff}$ one has to take into account both one-particle
irreducible (OPI) four-vertex graphs and one particle Higgs reducible
graphs with four external vector particles, as shown in Fig.~1. 
\begin{figure}
\setlength{\unitlength}{1cm}
\begin{picture}(12,5)
\put(2,0){\epsfig{file=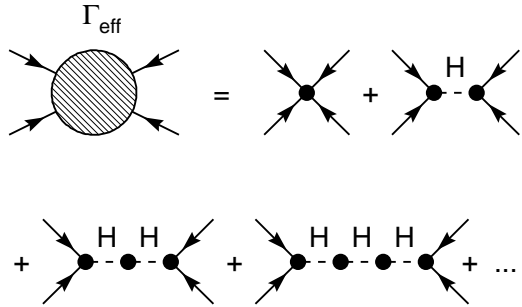,height=5cm}}
\end{picture}
\caption{One particle irreducible and Higgs reducible graphs
contributing to low energy quartic vector boson vertex. Bold blobs
denote the one particle irreducible four-, three- and two-point vertex
functions.}
\end{figure}

The two-loop topologies and one-loop topologies with counterterm
insertions contributing to OPI four-, three-, and
two-point vertex functions are shown in Fig.~2. The numbers in
parentheses show the total number of corresponding topologies, the
external lines are assumed to be topologically different.

\begin{figure}
\setlength{\unitlength}{1cm}
\begin{picture}(12,10)
\put(2,0){\epsfig{file=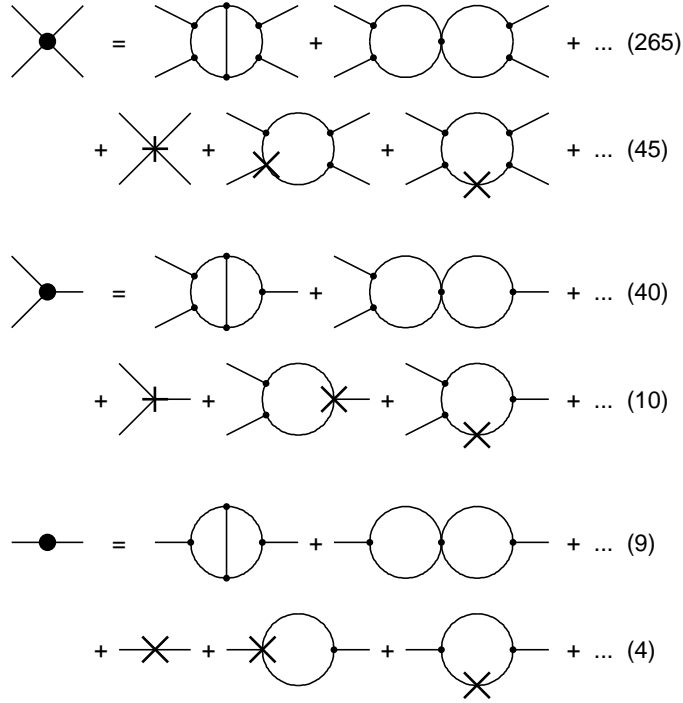,height=10cm}}
\end{picture}
\caption{One-particle irreducible two- and one-loop topologies.}
\end{figure}

Here we present only the simplest terms of the two-loop low-energy effective
action $\Gamma_{eff}$ to order $m_H^2$, which contribute to the
$ZZZZ$ anomalous quartic vertex, which is not present in the \SM{} at the
tree-level:
\begin{eqnarray}
&&\frac{\delta\Gamma_{eff}}{\delta Z_{\mu_1}(k_1)\delta Z_{\mu_2}(k_2)
\delta Z_{\mu_3}(k_3)\delta Z_{\mu_4}(k_4)} =  
\nonumber\\
&&\frac{e^6}{(16\pi^2)^2\sw^6\cw^4} \frac{\mH^2}{\mW^2} 
\biggl(g^{\mu_1 \mu_2} g^{\mu_3 \mu_4} + g^{\mu_1 \mu_3} g^{\mu_2 \mu_4} 
  + g^{\mu_1 \mu_4} g^{\mu_2 \mu_3}\biggr)
\label{ZZZZ}\\
&&\times \left(
          + \frac{337}{64}
          - \frac{39}{16} \pi Cl
          + \frac{105}{64} \pi \sqrt{3}
          - \frac{557}{576} \pi^2
          - 2 Cl \sqrt{3}
          + \frac{63}{16} \zeta(3)
          \right).
\nonumber
\end{eqnarray}

As is well known \cite{rho,pisa,masses,triple} the two-loop $m_H^2$
corrections to fermion scattering processes and triple vector boson
couplings are very small, in spite of the $m_H^2/\mW^2$ enhancements,
not only because of the small two-loop factor $g^4/(16\pi^2)^2$, but
also because the dimensionless coefficients themselves are of the
order of $10^{-1}-10^{-2}$, {\it i.e.} quite small . In this respect
the $W^+W^-W^+W^-$, $W^+W^-ZZ$ and $ZZZZ$ quartic couplings represent
a drastic contrast to the other vertices.  The dimensionless
coefficients in (\ref{ZZZZ}) are about 2,
{\it i.e.} about 20 times larger, than the largest dimensionless
coefficients for fermionic and triple vector boson couplings!  As was
mentioned previously, these particular vertices are distinguished, due
to a contribution from two-loop Higgs self energy insertion in the
Higgs-reducible graphs.  These vertices receive a contribution from
the $\zeta(3)$ and $\pi Cl$ terms, which originate only from the
two-loop Higgs mass counterterm \cite{our} as a term
proportional to a linear combination $21\zeta(3) -13\pi Cl$. In a
sense these couplings could be considered ``genuine'' quartic
couplings, which are the most sensitive to the details of the
mechanism of the electroweak symmetry breaking.

\section{Numerical results}

The possibilities to probe the quartic vector boson couplings through
the $WW$-, $ZZ$-fusion reactions 
\begin{eqnarray}
&&pp\to VVX,\label{LHC},\\
&&ee\to VVff,\label{ee-nnVV}
\end{eqnarray}
at the CERN Large Hadron Collider (LHC) or the electron-positron
linear collider are under intense study.  Here $V=\gamma$, $Z$ or
$W^{\pm}$ and $f=e$ or $\nu_e$.

In order to demonstrate the potential importance of large Higgs mass
corrections at high energies, we present in Fig.~3 the energy
dependence of the Born and corrected cross section of vector boson
scattering integrated over scattering angles in the region $30^\circ <
\theta < 150^\circ$ for $\sqrt{s_{VV}}$ up to 1~TeV for the very heavy
Higgs boson mass of 1.5~TeV. The existence of a physical Higgs
particle with such large mass seems to be excluded due to triviality
bounds (see \cite{LP'97-2} and references therein). We can consider
however such a value of the $m_H$ as an effective ultraviolet cut-off
in the theory without visible scalar Higgs particle. We see that the
growth with energy of the longitudinal vector boson scattering cross
sections, which is the experimental indication of the existence of
heavy Higgs sector and/or strong interactions among longitudinal
$W_L$, $Z_L$ bosons, is strongly modified by the two-loop $m_H^2$
corrections.  At high energy the cross sections of neutral channel
reactions are diminished, and those of charged channel reactions are
enhanced. The large value of the two-loop correction is not only due
to $m_H^2/M_W^2$ enhancement factor, but also due to violations of
unitarity cancellations in the presence of anomalous quartic
couplings. {\it E.g.} at 500~GeV correction to longitudinal $WW$
scattering for $m_H=1.5$~TeV is 
\begin{equation}
\delta \sim \frac{e^4}{(16\pi^2)^2\sw^4}\frac{m_H^2}{M_W^2}\times 
\frac{s}{M_W^2}\sim 0.25\%\times 40\sim 10\%,
\end{equation}
which gives correct order of magnitude for corrections in Fig.~3.

\begin{figure*}
\setlength{\unitlength}{1cm}
\begin{picture}(11.5,16.5)
\put(1,0){\epsfig{file=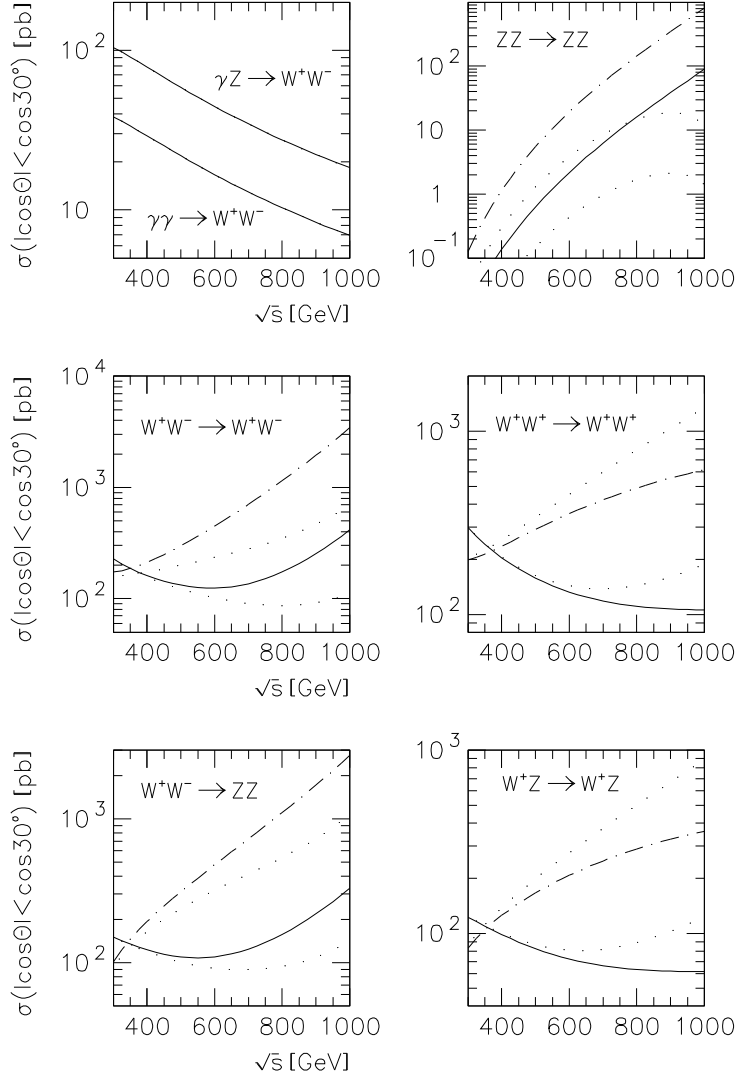,height=16.5cm}}
\end{picture}
\caption{Energy dependence of the cross sections for unpolarized
$UUUU$ (solid lines) and longitudinal $LLLL$ (dash-dotted lines)
vector boson scattering reactions. Dotted lines show corresponding
corrected cross sections. Higgs mass is taken to be 1.5~TeV}
\end{figure*}

Of course at center-of-mass energy of 1~TeV $s_{VV}$ is not very much
smaller than $m_H^2$, which is the condition under which our
low-energy effective action was calculated. Nevertheless, we think
that the qualitative trend, namely the fact that the account of large
Higgs mass corrections at high energy can change the value of the
cross section by a large factor of $2- 4$, is important for all
considerations of the signal from strong scattering of longitudinal
vector boson at TeV energy.

In fact, using the results of a thorough phenomenological analysis of
the effects of anomalous quartic couplings in $pp$ and $e^{\pm} e^-$
collisions \cite{oscar,ee-nnVV} we can estimate the potential of TeV
colliders in investigating the effects of enhanced $m_H^2$ two-loop
corrections more quantitatively. Anomalous quartic couplings are
defined in Refs. \cite{oscar,ee-nnVV} through the following effective
electroweak chiral Lagrangians: \small
\begin{eqnarray}
{\cal L}_4 &=& g^4\alpha_4\Biggl[\frac{1}{2}[(W^+W^-)^2+(W^{+2})(W^{-2})]
+\frac{1}{\cw^2}(W^+Z)(W^-Z)+\frac{1}{4\cw^4}Z^4\Biggr],\nonumber\\
{\cal L}_5 &=& g^4\alpha_5\Biggl[(W^+W^-)^2
+\frac{1}{\cw^2}(W^+W^-)Z^2+\frac{1}{4\cw^4}Z^4\Biggr],\label{L5}
\end{eqnarray}
\normalsize
where $g=e/\sw$. These operators introduce all possible quartic
couplings among the weak gauge bosons, that are compatible with
custodial $SU(2)_c$ symmetry \cite{chiral}. Although our complete
effective action given in \cite{our} does not obey this symmetry and
as a consequence can not be described by the combination of operators
(\ref{L5}), the dominating terms which originate from
two-loop Higgs self energy insertions in the Higgs reducible graphs
have exactly the structure of Lagrangian (\ref{L5}). Using our
expression (\ref{ZZZZ}) and analogous expressions for $WWWW$ and
$WWZZ$ vertices we can {\it calculate} the coupling constant
$\alpha_5$:
\begin{equation}
\alpha_5 \approx - \frac{g^2}{(16\pi^2)^2}\frac{m_H^2}{\mW^2}.
\label{a5}
\end{equation}
In our approach the constant $\alpha_4$ should be about an order of
magnitude smaller.

The potential of the LHC and TeV $e^\pm
e^-$ linear collider to study anomalous quartic vector boson
interactions was carefully analyzed in recent papers
\cite{ee-nnVV,oscar}. The limit on the anomalous quartic coupling
$\alpha_5$ which will be accessible at LHC is \cite{oscar}:
\begin{equation}
-7.2\leq \alpha_5 \leq 13.
\end{equation}
An integrated luminosity of 100$fb^{-1}$ was assumed. The invariant
mass of the vector boson pair was required to be in the range
$0.5<M_{VV}<1.25$~TeV. As one could expect from Fig.~3, the lower
bound on $\alpha_5$ is determined by the limits from same sign $W^\pm
W^\pm$-pair production, because for negative $\alpha_5$ (\ref{a5})
correction in this channel is positive.

Study of the reaction (\ref{ee-nnVV}) at the linear collider running at
1.6~TeV energy will be able to improve the LHC limits by a factor of
five \cite{ee-nnVV}.  Indeed, the 90\% bound on $\alpha_5$, obtained
by combining the $e^+e^-\to\nu_e\bar\nu_e W^+W^-$ and
$e^+e^-\to\nu_e\bar\nu_e ZZ$ channels is
\begin{equation}
|\alpha_5|\le 1.5\times 10^{-3}
\end{equation}
for integrated luminosity of 500~fb$^{-1}$. These limits were obtained
under the assumption that only anomalous parameter $\alpha_5$ is
non-vanishing.

For the Higgs mass of 1.5~TeV the value of $\alpha_5$ from
Eq. (\ref{a5}) is approximately $-6\times 10^{-3}$ (and $-2\times
10^{-3}$ for $m_H=900$~GeV), which is four times larger than the
experimental limit achievable at the linear collider. This comparison
is a very good indication that in the case, if a heavy Higgs scenario
of the electroweak symmetry breaking is realized in nature, its
indirect quantum effects could probably be measured.

\end{document}